\documentclass{article}
\usepackage{ltwol2e}

\def\Journal#1#2#3#4{{#1} {\bf #2}, #3 (#4)}

\def\NPB{{\em Nucl. Phys.} B}
\def\PLB{{\em Phys. Lett.}  B}
\def\PRL{\em Phys. Rev. Lett.}
\def\PRD{{\em Phys. Rev.} D}

\def\be{\begin{equation}}
\def\ee{\end{equation}}
\def\bea{\begin{eqnarray}}
\def\eea{\end{eqnarray}}

\begin{document}

\title{Flavor Asymmetry of the Sea Quarks in the Baryon Octet}

\author{Susumu Koretune}

\address{Department of Physics , Shimane Medical University , \\
Izumo,Shimane 693,Japan\\E-mail:koretune@shimane-med.ac.jp}   


\twocolumn[\maketitle\abstracts{We show that the chiral $SU(n)\otimes SU(n)$ flavor symmetry
on the null-plane severely restricts the 
sea quarks in the baryon octet. It predicts large
asymmetry for the light sea quarks $(u,d,s)$, and universality and 
abundance for the heavy sea quarks. Further it is shown
that existence of the heavy sea quarks constrained by the
same symmetry reduces the theoretical value of the 
Ellis-Jaffe sum rule substantially.}]
\section{Introduction}
Many years ago, based on the current anti-commutation relation
on the null-plane~\cite{K80}, the Gottfried sum rule 
~\cite{Got} was re-derived. Since the re-derived sum rule had a 
slightly different physical meaning from the original one,
I called it as the modified Gottfried
sum rule~\cite{K84}. Several years ago, this sum rule was
found to take the following form~\cite{K93,K95}:
\begin{equation}
\begin{array}{l}
\int^1_0\frac{dx}{x}\{F_2^{ep}(x,Q^2)-F_2^{en}(x,Q^2)\}=\frac{1}{3}( 1-\frac{4f_{K}^2}{\pi}\\
\times \int_{m_Km_N}^{\infty}\frac{d\nu}{\nu^2}
\sqrt{\nu^2-(m_Km_N)^2}\{\sigma^{K^+n}(\nu)-\sigma^{K^+p}(\nu)\}) ,
\end{array}\label{eq:mgott}
\end{equation}
where $\sigma^{K^+N}(\nu)$ with $N = p$ or $n$ is the total cross section of the 
$K^+N$ scatterings and $f_K$ is the kaon decay constant.
This gave us a new way to investigate the vacuum properties
of the hadron based on the chiral $SU(n)\otimes SU(n)$ 
flavor symmetry on the null-plane.  Here I briefly explain
the fact and show that it severely restricts the sea quarks
in the ${\bf 8}$ baryon. For details see Ref.~\cite{K98}.
\section {A physical meaning of the modified Gottfried sum rule}
The Gottfried sum multiplied by $3/2$ takes the following form:
\begin{equation}
\begin{array}{l}
\frac{3}{2}\int^1_0\frac{dx}{x}\{F_2^{ep}(x,Q^2)-F_2^{en}(x,Q^2)\}
=\int_0^1dx\{\frac{1}{2}u_v - \frac{1}{2}d_v\} \\
+\int_0^1dx\{\frac{1}{2}\lambda_u - \frac{1}{2}\lambda_d\}
- \int_0^1dx\{- \frac{1}{2}\lambda_{\bar{u}} + 
\frac{1}{2}\lambda_{\bar{d}}\}.
\end{array}\label{eq:gott}
\end{equation}
On the other hand Adler sum rule takes the following form:
\begin{equation}
\begin{array}{l}
\frac{1}{4}\int^1_0\frac{dx}{x}\{F_2^{\bar{\nu}p}(x,Q^2)-F_2^{\nu p}(x,Q^2)\}
=\int_0^1dx\{\frac{1}{2}u_v - \frac{1}{2}d_v\} \\+
\int_0^1dx\{\frac{1}{2}\lambda_u - \frac{1}{2}\lambda_d\}
+ \int_0^1dx\{- \frac{1}{2}\lambda_{\bar{u}} + 
\frac{1}{2}\lambda_{\bar{d}}\}.
\end{array}\label{eq:adler}
\end{equation}
Here the subscript $v$ means the valence quark and $\lambda_i$ means the 
$i$ type sea quark.
The fundamental difference between the Gottfried sum and the
Adler sum rule is the sign in front of the antiquark distribution
function. Thus, under the physically reasonable assumption 
$\int^1_0dx\lambda_i(x,Q^2)= \int^1_0dx\lambda_{\bar{i}}(x,Q^2)$,
we can say that the Adler sum rule measures
the mean $I_3$ of the \{[quark] + [antiquark]\} and hence the
one of the valence quarks being equal to the $I_3$ of the proton,
while the modified Gottfried sum rule measures the 
mean $I_3$ of the \{[quark] - [antiquark]\}in the proton.
Now the current commutation relation on the null-plane is given as
\begin{equation}
\begin{array}{l}
[J_a^+(x),J_b^+(0)]|_{x^+=0}=[J_a^{5+}(x),J_b^{5+}(0)]|_{x^+=0} \\
=if_{abc}\delta(x^-)\delta^2(\vec{x}^{\bot })J_c^+(0). 
\end{array}\label{eq:com}
\end{equation}
and the current anti-commutation relation on the null-plane~\cite{K80} is given as
\begin{equation}
\begin{array}{l}
<p|\{J_a^+(x),J_b^+(0)\}|p>_c|_{x^+=0} \\
= <p|\{J_a^{5+}(x),J_b^{5+}(0)\}|p>_c|_{x^+=0} \\
=\frac{1}{\pi }P(\frac{1}{x^-})\delta^2(\vec{x}^{\bot })
[d_{abc}A_c(p\cdot x , x^2=0) \\
\hspace*{2cm}+ f_{abc}S_c(p\cdot x , x^2=0)]p^+ .
\end{array}\label{eq:antcom}
\end{equation}
Both-hand sides of the modified Gottfried sum rule are equal to
\begin{equation}
\frac{1}{3\pi}P\int_{-\infty}^{\infty}\frac{d\alpha}{\alpha}
A_3(\alpha ,0),
\label{eq:a3}
\end{equation}
where as far as we are discussing the moment at $n=1$
$A_a(\alpha ,0)$can be related to the bilocal currents on the null-plane as
\begin{equation}
\begin{array}{l}
<p|\frac{1}{2i}[:\bar{q}(x)\gamma^{\mu}\frac{\lambda_a}{2}q(0):
- :\bar{q}(0)\gamma^{\mu}\frac{\lambda_a}{2}q(x):]|p>_c\\
=p^{\mu}A_a(p\cdot x , x^2) +x^{\mu}\bar{A}_a(p\cdot x , x^2).
\label{eq:bilocal}
\end{array}
\end{equation}
We decompose the quark field on the null-plane into the particle
mode and the anti-particle mode as
\begin{equation}
 q(x)=\sum_na_n\phi_n^{(+)}(x) +\sum_nb_n^{\dagger}\phi_n^{(-)}(x).
\end{equation}
The normal ordered product on the null-plane is
\begin{equation}
\begin{array}{l}
:\bar{q}(x)\gamma^+ \frac{\lambda_3}{2}q(0):
= \sum_{n,m}a_n^{\dagger}a_m\bar{\phi}_n^{(+)}(x)\gamma^+ 
\frac{\lambda_3}{2}\phi_m^{(+)}(0)\\
 - \sum_{n,m}b_m^{\dagger}
b_n\bar{\phi}_n^{(-)}(x)\gamma^+   
\frac{\lambda_3}{2}\phi_m^{(-)}(0) .
\end{array}
\end{equation}
By setting $\alpha =p^+y^-$,we define a part contributing to the quark
distribution function as
\begin{equation}
\begin{array}{l}
f(x)=\frac{1}{2\pi p^+}\int_{-\infty}^{\infty}d\alpha 
\exp [-ix\alpha ] \\
\times <p|\sum_{n,m}a_n^{\dagger}a_m\bar{\phi}_n^{(+)}
(y^-)\gamma^+ \frac{\lambda_3}{2}\phi_m^{(+)}(0)|p>_c ,
\end{array}\label{eq:quark}
\end{equation}
and a part contributing to the antiquark distribution function as
\begin{equation}
\begin{array}{l}
g(x)=-\frac{1}{2\pi p^+}\int_{-\infty}^{\infty}d\alpha 
\exp [-ix\alpha ]\\
\times <p|\sum_{n,m}b_n^{\dagger}b_m\bar{\phi}_m^{(-)}
(0)\gamma^+ \frac{\lambda_3}{2}\phi_n^{(-)}(y^-)|p>_c .
\end{array}\label{eq:antquark}
\end{equation}
By using the integral representation of the sign function
\begin{equation}
\epsilon (x)=\frac{1}{i\pi}P\int_{-\infty}^{\infty}
\frac{da}{a} \exp [iax] ,
\end{equation}
we obtain
\begin{equation}
\begin{array}{l}
\frac{1}{i\pi p^+}P\int_{-\infty}^{\infty}\frac{d\alpha}{\alpha} 
<p|\sum_{n,m}a_n^{\dagger}a_m\bar{\phi}_n^{(+)}
(y^-)\gamma^+ \frac{\lambda_3}{2}\phi_m^{(+)}(0)|p>_c \\
=\int_0^1dx\epsilon (x)f(x) =\int_0^1dxf(x),
\end{array}\label{eq:quarksign}
\end{equation}
\begin{equation}
\begin{array}{l}
\frac{1}{i\pi p^+}P\int_{-\infty}^{\infty}\frac{d\alpha}{\alpha}
<p|\sum_{n,m}b_m^{\dagger}b_n\bar{\phi}_n^{(-)}
(y^-)\gamma^+ \frac{\lambda_3}{2}\phi_m^{(-)}(0)|p>_c\\
=-\int_0^1dx\epsilon (-x)g(x) =\int_0^1dxg(x).
\end{array}\label{eq:antquarksign}
\end{equation}
The integral of the type $P\int_{-\infty}^{\infty}\frac{d\alpha}{\alpha}\cdots $ originates
from the factor $P\frac{1}{x^-}$ in Eq.~(\ref{eq:antcom}). 
In case of the commutation relation we obtain the integral
of the type $\int^{\infty}_{-\infty}d\alpha\delta(\alpha)\cdots $ which originates
from the factor $\delta (x^-)$in Eq.~(\ref{eq:com}), hence we have no sign chage as in  
Eq.~(\ref{eq:antquarksign}). Thus the current anti-commutation relation
on the null-plane naturally explains the physical difference
between the Gottfried sum and the Adler sum rule, and at the same
time, we can see that the physical importance of this relation
lies in the fact that it gives us information of the vacuum property
of the hadron.
\section{The regularization of the divergent sum rule}
Now the sum rule from the current anti-commutation relation
can be expected to diverge in general. Here we explain the method to regularize 
such divergence by taking the following sum rule.
\begin{equation}
\begin{array}{l}
\int^1_0\frac{dx}{x}F_2^{ep} \\
= \frac{1}{18\pi}
P\int^{\infty}_{-\infty}\frac{d\alpha}{\alpha}\{
2\sqrt{6}A_0(\alpha ,0) + 3A_3(\alpha ,0) +
\sqrt{3}A_8(\alpha ,0)\} .
\end{array}\label{eq:divsum}
\end{equation}
The left-hand side of this sum rule definitely diverges because of the
pomeron. We assume this component as the flavor singlet.
Then we can identify the divergent piece on the right-hand side
as that coming from $A_0(\alpha ,0)$. If the trajectory of the
pomeron takes the value $\alpha_P(t)<1$ at some $t$ near zero,
we can regularize the sum rule by the analytical continuation from
the non-forward direction, since in this case we can derive
the finite sum rule.~\cite{K84} The soft pomeron by Donnachi and
Landshoff~\cite{DL} is one example which makes it possible
to carry out the program easily. Now the assumption $\alpha_P(t) < 1$ 
for some small $t$ can not be satisfied by the hard pomeron based on
the fixed-coupling constant~\cite{hard}.
However, there are great efforts to improve the defect
of this pomeron~\cite{For}. The next-to-leading
corrections seems to suggest a substantial reduction of the
value of the intercept~\cite{Cia}.
The multiple scatterings of the pomeron gives us important
unitary corrections at low $x$~\cite{Mue}. Thus even in such
a perturbative approach there is a hope to satisfy the
assumption. Though we use the soft pomeron to explain the 
regularization, in view of the situation, 
we clarify the quantities which do not depend on the
assumed high energy behavior in the
following. Now we continue to discuss by the effective method
which gives the same results as those in the non-forward direction.
We take the leading high energy behavior of the $F_2^{ep}$
is given by the pomeron as $(\frac{1}{Q^2})^{\alpha_P(0)-1}
\beta_{ep}(Q^2,1-\alpha_P(0))(2\nu )^{\alpha_P(0)-1}$,
and assume it to be the flavor singlet. It should be noted
that what we assume here is only the high energy behavior
$(2\nu )^{\alpha_P(0)-1}$ and no assumption is made about the
$Q^2$ dependence, since all the unknown $Q^2$ dependence
is absorbed in $\beta_{ep}$. This also applies to the
scale factor in $2\nu$.  
We rewrite the left-hand side of the sum rule as
\begin{equation}
\begin{array}{l}
\int_0^1 \frac{dx}{x}F_2^{ep} \\
=\int_0^1 \frac{dx}{x}
\{F_2^{ep} - \beta_{ep}(Q^2,1-\alpha_P(0))x^{1-\alpha_P(0)}
\}\\
+ \int_0^1\ dx \beta_{ep}(Q^2,1-\alpha_P(0))x^{-\alpha_P(0)} ,
\end{array}
\end{equation}
and the right-hand side of it as
\begin{equation}
\begin{array}{l}
\frac{\sqrt{6}}{9\pi}P\int_{-\infty}^{\infty}\frac{d\alpha}{\alpha}
A_0(\alpha ,0)\\
= \frac{\sqrt{6}}{9\pi}P\int_{-\infty}^{\infty}
\frac{d\alpha}{\alpha}\{A_0(\alpha ,0) - f(\alpha)\} 
+\frac{\sqrt{6}}{9\pi}P\int_{-\infty}^{\infty}\frac{d\alpha}{\alpha}
f(\alpha ) .
\end{array}
\end{equation}
By setting $\alpha_P(0)=1+b-\epsilon$, we expand $\beta_{ep}$
as $\beta_{ep}^0(Q^2) - (\epsilon -b)\beta_{ep}^1(Q^2) + 
O((\epsilon - b)^2)$. The pole term as $\epsilon \to b$ should
be canceled out from both-hand sides of Eq.~(\ref{eq:divsum}) since the sum
rules are convergent for the arbitrary finite positive
$(\epsilon -b)$ which
corresponds to the small negative $t$ in the non-forward case, hence there
must exists $f(\alpha )$ such that the quantity 
$\frac{2\sqrt{6}}{9}\widetilde{f}(\alpha )
= (\frac{2\sqrt{6}}{9}\frac{1}{2\pi}P\int_{-\infty}^{\infty}
\frac{d\alpha}{\alpha}f(\alpha ) - \frac{\beta^0_{ep}}{\epsilon -b})$
becomes finite in the limit $\epsilon \to b$,
where $\beta_{ep}^0$ is $Q^2$ independent since Eq.~(\ref{eq:divsum}) holds at any
$Q^2$. After taking out the singular piece we take the limit 
$\epsilon \to 0$ and obtain 
\begin{equation}
\begin{array}{l}
\int^1_0\frac{dx}{x}\{F_2^{ep}-\beta_{ep}^0x^{-b}\}\\
=\frac{1}{18\pi}
P\int^{\infty}_{-\infty}\frac{d\alpha}{\alpha}\{
2\sqrt{6}S_0^3(\alpha ,0,Q^2) + 3A_3^p(\alpha ,0) +
\sqrt{3}A_8^p(\alpha ,0)\} ,
\end{array}
\end{equation}
where $S_0^3(\alpha ,0,Q^2)$ and $\widetilde{S}_0^3$ are defined as
\begin{equation}
\begin{array}{l}
\frac{2\sqrt{6}}{9}\widetilde{S}_0^3\\
= \frac{2\sqrt{6}}{9}[\widetilde{f}(\alpha ) + 
\frac{1}{2\pi}P\int_{-\infty}^{\infty}\frac{d\alpha}{\alpha}
\{{A}_0(\alpha ,0)-f(\alpha)\}] + \beta_{ep}^1(Q^2) \\ 
= \frac{2\sqrt{6}}{9}\frac{1}{2\pi}P\int_{-\infty}^{\infty}
\frac{d\alpha}{\alpha}S_0^3(\alpha ,0,Q^2) .
\end{array}
\end{equation}
Here the superscript 3 in $S_0^3(\alpha ,0,Q^2)$ and
$\widetilde{S}_0^3$ means the singlet in the $SU(3)$.
We find that the regularization of the sum rule simply results in 
the $Q^2$ dependence in the singlet component 
and that all the relation from the symmetry is inherited. 
Thus if we get the relation which do not include the singlet
component, we can have regularization independent relation.
The modified Gottfried sum rule belongs to this type of the sum rules
and we give such sum rule in the following.
\section{The symmetry constraint on the light sea quark distributions
in the baryon octet}
$A_a(\alpha ,0)$ is governed by the 
chiral $SU(3)\otimes SU(3)$ flavor symmetry. 
If we take the state 
as the ${\bf 8}$ baryon, the matrix element becomes
\begin{equation}
\begin{array}{l}
\langle \alpha , p|\frac{1}{2i}[:\bar{q}(x)\gamma^{\mu}\frac{1}{2}\lambda_a q(0)
-\bar{q}(0)\gamma^{\mu}\frac{1}{2}\lambda_a q(x):]|\beta ,p\rangle_c\\
=p^{\mu}(A_a(px,x^2))_{\alpha \beta }+x^{\mu}(\bar{A}_a(px,x^2))_{\alpha \beta },
\end{array}
\end{equation}
where the $\alpha , \beta $ are the symmetry index 
specifying each member of the ${\bf 8}$ baryon.  
Since the matrix element can be classified by the flavor singlet in the
product ${\bf 8\otimes 8\otimes 8}$, $(A_a(\alpha ,0))_{\alpha  \beta }$ 
is decomposed as
\begin{equation}
(A_a(\alpha , 0))_{\alpha  \beta } = if_{\alpha a \beta}F(\alpha,0)
+ d_{\alpha  a \beta }D(\alpha,0)
\end{equation}
for $a\neq 0$ .  Using the value of the modified Gottfried
sum rule, we obtain~\cite{K93}
\begin{equation}
\begin{array}{l}
\frac{1}{3\pi}P\int_{-\infty}^{\infty}\frac{d\alpha}{\alpha}
A_3^p(\alpha ,0)\\
 =  \frac{1}{3\pi}\int_{-\infty}^{\infty}\frac{d\alpha}{\alpha}
\{ F(\alpha,0) + D(\alpha , 0)\}= 0.26 . 
\end{array}
\end{equation}
The mean hypercharge sum rule~\cite{K97} gives us
\begin{equation}
\begin{array}{l}
\frac{\sqrt{3}}{3\pi}P\int_{-\infty}^{\infty}\frac{d\alpha}{\alpha}
A_8^p(\alpha ,0)\\
= \frac{\sqrt{3}}{3\pi}\int_{-\infty}^{\infty}\frac{d\alpha}{\alpha}
\sqrt{3}\{ F(\alpha,0) - \frac{1}{3}D(\alpha , 0)\} = 2.12 .
\end{array}
\end{equation}
Here we use the notation $A_a^B(\alpha ,0)$ with 
$B = ( p , n , \Sigma^{\pm} ,\Sigma^0 , \Lambda^0 ,
\Xi^- , \Xi^0 )$ to specify each member of the ${\bf 8}$
baryon. Thus we obtain 
\begin{equation}
\widetilde{F} \equiv \frac{1}{2\pi}\int^{\infty}_{-\infty}
\frac{d\alpha}{\alpha}F(\alpha ,0) = 0.89 ,
\end{equation}
\begin{equation}
\widetilde{D} \equiv \frac{1}{2\pi}\int^{\infty}_{-\infty}
\frac{d\alpha}{\alpha}D(\alpha ,0) = -0.50 .
\end{equation}
Now we define the sea quark distribution 
of the $i$ type one in the ${\bf 8}$ baryon as $\lambda^B_i$, and
regularizes its mean number as
\begin{equation}
\langle \widetilde{\lambda}^B_i\rangle = \int_0^1dx \{
\lambda_i^B - ax^{-\alpha_P(0)}\} ,
\label{eq:regsea}
\end{equation}
where $\alpha_P(0)$ here is $\alpha_P(0)=1+b$ and is taken as
$1.0808 $, and $a$ is defined as
\begin{equation}
\lim_{x\to 0}x^{\alpha_P(0)}\lambda_i^B =a .
\end{equation}
The value $a$ is proportional to $\beta_{ep}^0$ and was 
determined through the sum rule as $a=1.2$.~\cite{K95}
Though this relation can be lost except the soft
pomeron case, we still have the fact that the singlet component
contributes universally to all the sea quarks.
Thus similar regularized sea quark number as Eq.~(\ref{eq:regsea})
exists in the general case.
The constraints on the sea quark distributions are 
obtained as follows. 
Let us take an example of the proton matrix element 
with $K=\frac{\lambda_a}{2} = 
diag(1\; 0\; 0)=\frac{\sqrt{6}}{6}\lambda_0
+ \frac{1}{2}\lambda_3 + \frac{\sqrt{3}}{6}\lambda_8$ .
Since $\alpha = \frac{1}{2}(4 + i5) ,
\beta = \frac{1}{2}(4 - i5) $, by taking 
$<\widetilde{\lambda}^B_i> = <\widetilde{\lambda}^B_{\bar{i}}>$,
we obtain $<u_v^p> + 2<\widetilde{\lambda}^p_u> =
\frac{\sqrt{6}}{3}\widetilde{S}^3_0 + 2\widetilde{F}
+ \frac{2}{3}\widetilde{D}$.
Thus we have many relations for the sea quarks in the
${\bf 8}$ baryon. 
\begin{table*}
\caption{{ The mean quantum number of the light sea quarks.}
}
\begin{center}
\begin{tabular}{|l|l|l|l|}\hline
 &$\langle I_3 \rangle$
 &$\langle Y \rangle$
 &$\langle Q \rangle$\\ 
$B$ &$\frac{1}{2}\{\langle \lambda_u^B - \lambda_d^B\rangle \}$ 
 &$\frac{1}{3}\{\langle \lambda_u^B + \lambda_d^B -2\lambda_s^B\rangle \}$
 &$\frac{1}{3}\{\langle 2\lambda_u^B - \lambda_d^B - \lambda_s^B\rangle \}$\\ \hline\hline 
&&& \vspace{-4mm}\\
$p$ 
 &$\frac{1}{2}(\widetilde{F} + \widetilde{D}) - \frac{1}{4}$
 &$\frac{1}{3}(3\widetilde{F} - \widetilde{D}) - \frac{1}{2}$
 &$\frac{1}{3}(3\widetilde{F} + \widetilde{D}) - \frac{1}{2}$ \\ 
 &$=-0.055$ &$ =0.56$ &$ =0.23$ \\ \hline 
$n$ 
 &$-\frac{1}{2}(\widetilde{F} + \widetilde{D}) + \frac{1}{4}$
 &$\frac{1}{3}(3\widetilde{F} - \widetilde{D}) - \frac{1}{2}$
 &$-\frac{2}{3}\widetilde{D} $ \\
 &$=0.055$ & $=0.56$ & $=0.34$ \\ \hline  
$\Sigma^+$ 
 &$\frac{1}{2}\widetilde{F} - \frac{1}{2}$
 &$\frac{2}{3}\widetilde{D} $
 &$\frac{1}{6}(3\widetilde{F} + 2\widetilde{D}) - \frac{1}{2}$  \\
 &$=-0.054$ & $=-0.34$ & $=-0.22$ \\  \hline 
$\Sigma^0$ 
 &$0$
 &$\frac{2}{3}\widetilde{D} $
 &$\frac{1}{3}\widetilde{D}$ \\
 &$=0 $& $=-0.34$ &$ =-0.17$ \\ \hline 
$\Sigma^-$ 
 &$-\frac{1}{2}\widetilde{F} + \frac{1}{2}$
 &$\frac{2}{3}\widetilde{D}$ 
 &$\frac{1}{6}(-3\widetilde{F} + 2\widetilde{D}) + \frac{1}{2}$ \\  
 &$=0.054$ & $=-0.34$ &$ =-0.11$ \\ \hline 
$\Xi^-$ 
 &$\frac{1}{2}(-\widetilde{F} + \widetilde{D}) + \frac{1}{4}$
 &$-\frac{1}{3}(3\widetilde{F} + \widetilde{D}) + \frac{1}{2}$
 &$-\frac{1}{3}(3\widetilde{F} - \widetilde{D}) + \frac{1}{2} $\\ 
 &$=-0.45$ & $=-0.23$ & $=-0.56$ \\ \hline 
$\Xi^0$ 
 &$\frac{1}{2}(\widetilde{F} - \widetilde{D}) - \frac{1}{4}$
 &$-\frac{1}{3}(3\widetilde{F} + \widetilde{D}) + \frac{1}{2}$
 &$-\frac{2}{3}\widetilde{D} $ \\ 
 &$=0.45$ & $=-0.23$ & $=0.34 $\\\hline 
$\Lambda^0 $
 &$0$
 &$-\frac{2}{3}\widetilde{D}$
 &$-\frac{1}{3}\widetilde{D}$ \\
 &$=0$ & $=0.34$ &$ =0.17$ \\\hline 
\end{tabular}
\end{center}
\end{table*}
Then by using the fact that each valence part is merely the 
number of the valence quark, we get many sum rules from 
these relations.
Among them the sum rules for the mean quantum numbers of the light sea
quarks are fundamental since they do not depend on 
the singlet component. In other words they do not depend on the
regularization. Here the light sea quarks
mean the $u,d,s$ type sea quarks . We summarize them in the Table. 
Note that $\widetilde{\lambda}^B_i$ is replaced by
$\lambda^B_i$ because the divergent part is canceled out
in each expression. The perturbative QCD corrections to
these relations begin from the 2 loops and 
they enter the same way as the one in the  modified
Gottfried sum rule.~\cite{Ross,CFP} Therefore they are negligibly small
compared with the non-perturbative values listed
in the Table. 
\section{The symmetry constraint on the heavy sea quark distributions
in the baryon octet}
Here we extend the symmetry from the $SU(3)\otimes SU(3)$
to the $SU(n)\otimes SU(n)$ with $n \ge 4$.  
Let us now discuss the heavy sea quarks in the ${\bf 8 }$
baryon. For concreteness we take chiral $SU(4)\otimes SU(4)$ 
flavor symmetry. In this case the ${\bf 8 }$ baryon
belongs to the ${\bf 20_M}$, and the currents to the
${\bf 15}$. The matrix element in this case can be
classified by the singlet component in the product
${\bf \bar{20}_M \otimes 20_M \otimes 15}$.
Since the adjoint representation ${\bf 15}$ 
appears twice in the product as \\
${\bf \bar{20}_M \otimes 20_M 
= 175 \oplus 84 \oplus 45 \oplus \bar{45} \oplus 20 \oplus
15 \oplus 15 \oplus 1}$,  and since only these two ${\bf 15}$ can 
make the singlet with the remaining ${\bf 15}$,
we have only two different terms in the matrix element.
Further these two ${\bf 15}$ can be represented
by the $4 \times 4$ matrix whose $3 \times 3$
sub-matrix agrees with the $3 \times 3$ matrix in the
$SU(3)$. Thus the two different terms are
$F(\alpha , 0)$ and $D(\alpha , 0)$ in the $SU(3)$. 
However , in this generalization,
the singlet in the $SU(3)$ is not the singlet
in the $SU(4)$. To see this fact more concretely, we
take the matrix $K = diag(1\; 0\; 0\; 0)$, 
and decomposes it as $K = \frac{\sqrt{2}}{4}\lambda_0^4
+ \frac{1}{2}\lambda_3^4 +\frac{\sqrt{3}}{6}\lambda_8^4
+ \frac{\sqrt{6}}{12}\lambda_{15}^4$. Here the $\lambda_k^4$ 
is the Gell-Mann matrix generalized to the
$SU(4)$. The $SU(3)$
singlet part in this decomposition is
$\frac{\sqrt{2}}{4}\lambda_0^4 + \frac{\sqrt{6}}{12}\lambda_{15}^4
= \frac{1}{3}diag(1\; 1\; 1\; 0)$ . Since the 
$3\times 3$ sub-matrix $diag(1\; 1\; 1)$ is expressed as
$\frac{\sqrt{6}}{2}\lambda_0$ in the $SU(3)$, the coefficient
of the singlet part
in the $SU(3)$ is different from the one in the $SU(4)$.
On the other hand, $3\times 3$ sub-matrix in the part
$\frac{1}{2}\lambda_3^4 + \frac{\sqrt{3}}{6}\lambda_8^4$
has the same expression in these two cases. Thus we find
one relation between the singlet contribution in the $SU(3)$
and the one in the $SU(4)$. If we denote the $SU(4)$ singlet 
contribution as $\widetilde{S}^4_0$ corresponding to the 
$\widetilde{S}^3_0$ in the $SU(3)$, we obtain 
$\frac{\sqrt{6}}{3}\widetilde{S}^3_0
= \frac{\sqrt{2}}{2}\widetilde{S}^4_0 + \frac{1}{3}\widetilde{D}$.
Expressed in the parton model,
this generalization from the $SU(3)$ to the $SU(4)$ corresponds
to the addition of the charm sea quark with 
the symmetric condition without changing
anything in the light sea quarks. Thus we see
all the charm sea quark distributions in the
${\bf 8}$ baryon corresponding
to the matrix element of $diag(0\; 0\; 0\; 1)$ are the
same. Explicitly we obtain $2<\widetilde{\lambda}_c> =
\frac{\sqrt{6}}{3}\widetilde{S}^3_0 - \frac{4}{3}\widetilde{D}$ for all  
members in the ${\bf 8}$ baryon. For the proton, we obtain
$<\lambda_c - \lambda_d^p>=0.5$ and
$<\lambda_c - \lambda_s^p>=1.4$. This means that the charm sea quark 
in the proton is abundant especially in the small $x$ region.~\cite{K95,K92}
It should be noted that the gluon 
fusion like term is in general included in our definition of the 
charm quark distribution function, since the virtual photon couples to
the gluon through the quark and since the classification of the sea
quark in our case is done by this coupling with the virtual photon.
Hence we reach the conclusion
that the charm sea quark is universal and abundant in 
the ${\bf 8}$ baryon. \\
The same kind of the discussions can be repeated in the 
$SU(5)$ or $SU(6)$, and we get 
$ 2<\widetilde{\lambda}_b> =2<\widetilde{\lambda}_t> 
=\frac{\sqrt{6}}{3}\widetilde{S}^3_0 - \frac{4}{3}\widetilde{D} $ 
for the bottom and the top sea quarks.
\section{Flavor asymmetry of the spin-dependent sea quark
distribution}
It is interesting to note that similar discussion
to extend the symmetry from the $SU(3)$ to the
$SU(4)$ can be applied to the matrix element
$<p,s,\alpha | J^{5\mu}_a(0)|p,s,\beta >$.  We define
\begin{equation}
\langle p,s,\alpha |J_a^{5\mu}(0)|p,s,\beta \rangle = s^{\mu}A_a^{\alpha
\beta}  ,
\end{equation}
where $s^{\mu}$ is the spin vector, and 
\begin{equation}
A_a^{\alpha \beta }=if_{\alpha a \beta }F + d_{\alpha a \beta }D ,
\end{equation}
for $a \neq 0$. The Ellis-Jaffe sum~\cite{Ellis} for the ${\bf 8}$ baryon is
\begin{equation}
I_f^B=\int_0^1 dxg_1^{B}(x,Q^2) ,
\end{equation}
where the subscript $f$ specifies the flavor group.
$I_f^B$ is proportional to $d_{\alpha a \beta}$ and in case
of the proton it is well known to take the form 
\begin{equation}
I_3^p = \frac{1}{36}[4\triangle Q_0^p + 3\triangle Q_3^p
+\triangle Q_8^p ] ,
\end{equation}
where $\triangle Q_0^p =\triangle u^p + \triangle d^p + \triangle s^p $,
$\triangle Q_3^p=\triangle u^p - \triangle d^p$, and $\triangle Q_8^p = 
\triangle u^p + \triangle d^p - 2\triangle s^p $, and $\triangle q^p$
is the fraction of the spin of the proton carried by the spin
of quarks of flavor $q$. Here $\triangle q^p$ includes the 
contribution from the antiquark as usual.
We obtain 
\begin{equation}
\triangle u^p=\frac{1}{3}S + \frac{1}{3}D +F ,
\triangle d^p=\frac{1}{3}S - \frac{2}{3}D ,
\triangle s^p=\frac{1}{3}S + \frac{1}{3}D - F, 
\end{equation}
where $\triangle Q_0^p=S$.  It is straightforward to get
the spin fraction of the quarks in other baryons.
Now in  the $SU(4)$, $\triangle Q_{15}^B$ can be defined as
\begin{equation}
\triangle Q_{15}^B=\sqrt{6}A_{15}^{\alpha \beta}=\triangle u^B 
+ \triangle d^B + \triangle s^B - 3\triangle c^B.
\end{equation}
Applying similar discussion in the previous section to this quantity we obtain 
$\triangle Q_{15}^B = 2D$ for all members in the ${\bf 8 }$ baryon.
Since $\triangle Q_{a}^B$ for $1\le a \le 8$ is the same as 
in  the $SU(3)$, we obtain 
\begin{equation}
\triangle c=\frac{1}{3}S - \frac{2}{3}D ,
\end{equation}
for all members in the ${\bf 8 }$ baryon. Note that we use the same
$S$ as in the $SU(3)$. Thus we get
\begin{equation}
\begin{array}{l}
I_4^p=\frac{1}{2}[\frac{4}{9}\triangle u^p 
+ \frac{1}{9}\triangle d^p + \frac{1}{9}\triangle s^p
+ \frac{4}{9}\triangle c^p ]\\
= \frac{5}{27}S + \frac{1}{6}F -\frac{5}{54}D .
\end{array}
\end{equation}
The generalization to the $SU(5)$ or the $SU(6)$ is 
straightforward, and we obtain
\begin{equation}
\triangle b = \triangle t = \frac{1}{3}S - \frac{2}{3}D .
\end{equation}
Using experimental value of $F=0.46\pm 0.01$ and $D=0.79\pm 0.01$~\cite{Hsu},
 we see that for a reasonable value of the $S$,
the theoretical value of the Ellis-Jaffe sum rule is reduced
substantially by the charm quark.  It is usually considered that
the light sea quark gets contribution from the
gluon anomaly because of the small-ness of the quark mass
compared with the infra-red cutoff~\cite{Mank}.
The magnitude of this gluon contribution is determined 
by input information. Then, to make the Ellis-Jaffe
sum rule consistent with the experiment by
this gluon polarization, it must be taken very large.
In our case, such large gluon polarization is not
necessary.  The heavy quarks such as the charm 
and the bottom ones
are suffice to make the Ellis-Jaffe sum rule
consistent with the experiment.    
 
\section{Conclusions}
We show that the chiral
$SU(n)\otimes SU(n)$ flavor symmetry on the null-plane
combined with the fixed-mass sum rule derived
from the current anti-commutation relation on the
null-plane  severely
restricts the sea quark in the ${\bf 8}$
baryon. It predicts a large asymmetry for the
light sea quarks, and universality and abundance
for the heavy sea quarks. 
Further we show
that the same symmetry restricts the fraction of the spin 
of the ${\bf 8}$ baryon carried by the quark.
Especially we show that 
this effect is outstanding for the intrinsic 
charm sea quark in the nucleon and that it plays the 
role to reduce the theoretical value of 
the Ellis-Jaffe sum rule substantially.


\section*{References}
\begin{thebibliography}{30}
\bibitem{K80}S.Koretune,  \Journal{\PRD}{21}{820}{1980}.
\bibitem{Got}K.Gottfried, \Journal{\PRL}{19}{1174}{1967}.
\bibitem{K84}S.Koretune,  {$ Prog.\; \, Theor.\;\,  Phys.\;$} {\bf 72}, 821 (1984).
\bibitem{K93}S.Koretune,  \Journal{\PRD}{47}{2690}{1993}.
\bibitem{K95}S.Koretune,  \Journal{\PRD}{52}{44}{1995}.
\bibitem{K98}S.Koretune, hep-ph9804408, to be published in the Nuclear Physics B.
\bibitem{DL}A.Donnachie and P.V.Landshoff, 
\Journal{\PLB}{296}{227}{1992}.
\bibitem{hard}E.A.Kuraev, L.N.Lipatov, and V.S.Fadin,
{$Sov.\; \, Phys.\; \, JETP\;$} {\bf 45}, 199 (1977);\\
Ya.Ya.Balitsky and L.N.Lipatov,
{$Sov.\; \, J.\; \, Nucl.\; \, Phys.\;$} {\bf 28}, 822 (1978).
\bibitem{For}J.R.Forshaw and D.A.Ross, {\it Quantum Chromodynamics
and the Pomeron}\\ (Cambridge Univ. Press,Cambridge,UK,1997),\\
pp.113-138 and pp.204-237.
\bibitem{Cia}M.Ciafaloni and G.Camici, hep-ph9803389.\\
V.S.Fadin and L.N.Lipatov, hep-ph9802290.
\bibitem{Mue}A.H.Mueller, hep-ph9710531.
\bibitem{K97}S.Koretune,  {$ Prog.\; \, Theor.\;\,  Phys.\;$} {\bf 98}, 749 (1997).
\bibitem{Ross}D.A.Ross and C.T.Sachrajda, \Journal{\NPB}{149}{497}{1978},
\bibitem{CFP}G.Curci, W.Furmanski, and R.Petronzio,
\Journal{\NPB}{175}{127}{1980}.
\bibitem{K92}S.Koretune,  {$ Prog.\; \, Theor.\;\,  Phys.\;$} {\bf 88}, 63 (1992).
\bibitem{Ellis}J.Ellis and R.L.Jaffe,  \Journal{\PRD}{9}{1444}
{1974};ED {\bf 10}1669(1974).
\bibitem{Hsu}S.Y.Hsueh et al., \Journal{\PRD}{38}{2056}{1988}.
\bibitem{Mank}Mankiewicz and Sch\"afer., \Journal{\PLB}{242}{455}
{1990}; and references cited therein. 
\end{thebibliography}
\end{document}